\documentclass[12pt]{article}
\usepackage{graphicx}
\newcommand{\be}[1]{\begin{equation} \label{#1}}
\newcommand{\ee}{\end{equation}}

\newcommand{\bea}[1]{\begin{eqnarray} \label{#1}}
\newcommand{\eea}{\end{eqnarray}}

\newcommand{\refeq}[1]{(\ref{#1})}

\begin{document}
\begin{center}
{\Large \bf
%\title{
        Exact results on spin dynamics and
        multiple quantum dynamics in alternating
        spin-$\frac{1}{2}$ chains with
        $XY$-Hamiltonian at high temperatures.
      }
\end{center}
\vskip 5mm
\begin{center}
{\large \bf
%\author{
        E.B. Fel'dman  and  M.G. Rudavets
% \footnote{ To whom correspondence should be addressed. e-mail: feldman@icp.ac.ru}
       }
\end{center}
\vskip 5mm
%\address{
\begin{center}
%\affiliation{
       Institute of Problems of Chemical Physics, \\
       Russian Academy of Sciences, 142432 Chernogolovka,
       Moscow Region, Russia
\end{center}
\date{}

%%%%%%%%%%%%%%%%%%%%%%%%%%%%%%%%%%%%%%%%%%%%%%%%%%%%%%%%%%%%%%%%%%%%%%%%%%%%%%%%%%%%%%%%%%%%%%%

\abstract{
We extend the picture of a transfer
of nuclear spin-$\frac{1}{2}$
polarization along a
homogeneous one-dimensional chain with the
$XY$-Hamiltonian
to the inhomogeneous chain with alternating nearest neighbour
couplings and alternating Larmor frequencies. To this end, we
calculate exactly the spectrum of the spin-$\frac{1}{2}$ $XY$-Hamiltonian
of the alternating chain with an odd number of sites.
The exact spectrum  of the $XY$-Hamiltonian is also applied to study
the multiple quantum (MQ) NMR dynamics of the alternating spin-$\frac{1}{2}$
chain. MQ NMR spectra are shown to have the MQ coherences of zero and $\pm$ second orders
just as in the case of a homogeneous chain. The intensities of the MQ coherences
are calculated.
}
%\begin{abstract}
\nopagebreak


%%%%%%%%%%%%%%%%%%%%%%%%%%%%%%%%%%%%%%%%%%%%%%%%%%%%%%%%%%%%%%%%%%%%%%%%%%%%%

\section{Introduction}
The discovery of the exact solution of
spin-$\frac{1}{2}$
homogeneous one-dimensional chains with
the $XY$-Hamiltonian
\cite{LSM}, \cite{CG} gives
the observable and  measurable features unraveling the NMR
dynamics of
spin-$\frac{1}{2}$
homogeneous one-dimensional chains \cite{FBE}.
Although most of the NMR experiments for
one-dimensional spin chains
appear to be well-explained by means of the nuclear spin dynamics on
the homogeneous spin chains,
NMR spin dynamics beyond
the homogeneous chains
has attracted the attention recently. For example,
the experiments \cite{MBSBE} demonstrate the propagation of spin
wave excitations along
the inhomogeneous spin chains,
the mesoscopic  echo  has been observed \cite{PL}
due to the reflections of the spin waves at the boundaries of the
chain.

The presented paper is aimed at exploring the key differences
of the NMR of
the homogeneous spin-$\frac{1}{2}$ chain
from the
inhomogeneous spin-$\frac{1}{2}$ chain. It may be difficult, however, to calculate
the NMR responses for
an inhomogeneous spin chain
with a random variation of the nearest neighbour (NN) dipolar coupling.
Thus, as a first step to unravel  of the
inhomogeneous effects, we treat the
one-dimensional chain
with an alternating
spin-$\frac{1}{2}$
NN dipolar coupling and alternating Larmor frequencies.

In the approach \cite{LSM}, \cite{CG}, the basic tool in  exploring
the $1D$ spin-$\frac{1}{2}$ $XY$-Hamiltonian is the Jordan-Wigner
transformation of the original spin-$\frac{1}{2}$ $XY$-Hamiltonian to the
Hamiltonian of the free fermions. In this way,
the exact thermodynamics for alternating infinite chains
with spin-$\frac{1}{2}$ the $XY$-Hamiltonian was explored in \cite{DR}.
The report \cite{YDX}
presents the exact spectrum of the $XY$-Hamiltonian with alternating couplings
on the finite rings.
However, to our knowledge, the spectrum
of the $XY$-Hamiltonian with alternating couplings on the open chains
is lacking by now. This is what will be addressed in section $2$.
The derived exact spectrum of the spin-$\frac{1}{2}$ $XY$-Hamiltonian of the
alternating open chains permits one to explain the
transfer of the nuclear polarization along the alternating chains
in section $3$ and to calculate the MQ intensities
of alternating spin-$\frac{1}{2}$  chains
in section $4$. The concluding section $5$ draws
the distinction of the NMR dynamics on alternating chains
from that on the homogeneous ones.

%%%%%%%%%%%%%%%%%%%%%%%%%%%%%%%%%%%%%%%%%%%%%%%%%%%%%%%%%%%%%%%%%%%%%%%%%%%%%

\section{Exact spectrum of spin-$\frac{1}{2}$  $XY$-Hamiltonian
               with alternating couplings on open chains
        }

\begin{figure}[ht]
\begin{center}
\includegraphics[angle=270,scale=0.6]{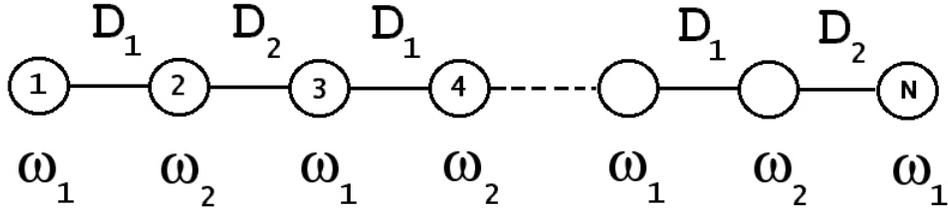}
\caption{
{\small
\label{Fig1}
An open chain of the odd number, $N$, of spins
with the alternating NN coupling constants $D_1$ and $D_2$
and alternating Larmor frequencies $\omega_1$, $\omega_2$.
}
        }

\end{center}
\end{figure}

In this section, we derive the exact spectrum of
the spin-$\frac{1}{2}$  $XY$-Hamiltonian
\bea{1}
H =
 \sum\limits_{n=1}^{N} \omega_n I_{nz}
+
\sum\limits_{n=1}^{N-1}
D_{n,n+1} \Bigl(
                 I_{n,x} I_{n+1,x} + I_{n,y} I_{n+1,y}
          \Bigr)
\eea
of the
open chain of the odd number, $N$, of sites with
the alternating NN coupling constants
$D_1$ and $D_2$
and alternating Larmor frequencies
$\omega_1$ and $\omega_2$, see Fig. $1$.

The nuclear spins are
specified by the
spin-$\frac{1}{2}$ operators $I_{n\alpha}$ at sites
$n = 1, \ldots, N$ with projections $\alpha = x, y, z$.
The Jordan-Wigner transformation \cite{LSM}
\bea{3}
I_{n,-} &=& I_{n,x} - i I_{n,y} = (-2)^{n-1}
  \left( \prod_{l=1}^{l=n-1} I_{l,z}
 \right) c_n, \nonumber \\
I_{n,+} &=& I_{n,x} + i I_{n,y} = (-2)^{n-1}
  \left( \prod_{l=1}^{l=n-1} I_{l,z}
 \right) c_n^{+}, \nonumber \\
I_{n,z} &=& c_n^{+}c_n - \frac{1}{2},
\eea
from spin-$\frac{1}{2}$ operators $I_{n\alpha}$ to the creation
(annihilation) operators $c_n^{+}$ ($c_n$) of the spinless
fermions takes the Hamiltonian \refeq{1} into the Hamiltonian
\bea{4}
H =
 \sum\limits_{n=1}^{N} \omega_n \left( c_n^{+}c_n - \frac{1}{2} \right)
+
\frac{1}{2}\sum\limits_{ n=1}^{N-1}
D_{n,n+1} \left\{
c_{n}^{+}c_{n+1} + c_{n+1}^{+}c_{n}
\right\},
\eea
or in the matrix notations as
\bea{5}
H =
\frac{1}{2} \overrightarrow{ c^{+}} \left( D + 2\Omega \right)
\overrightarrow{c} \: -
\:
\frac{1}{2} \sum\limits_{n=1}^{N} \omega_n.
\eea
In Eq. \refeq{5}, we denote the row vector
$\overrightarrow{ c^{+}} = ( c_{1}^{+},\ldots,c_{N}^{+}  )$,
the column vector
$\overrightarrow{ c} = ( c_{1},\ldots,c_{N} )^{t}$
 (the superscript $t$ represents the transpose) and specify
the matrices $\Omega$ and $D$ as
\bea{5.1}
\Omega=
\left[
\begin{array}{llllll}
\omega_1 &     0      &        0       &  \ldots      &   0         & 0           \\
  0      & \omega_2   &        0       &  \ldots      &   0         & 0           \\
  0      &     0      &    \omega_1    &  \ldots      &   0         & 0           \\
  \vdots &   \vdots   &    \vdots      & \vdots       &  \vdots     & \vdots      \\
  0      &     0      &        0       &  \ldots      &  \omega_2   & 0           \\
  0      &     0      &        0       &  \ldots      &   0         & \omega_1
\end{array}
\right], \quad
D=
\left[
\begin{array}{llllll}
  0      & D_1        &        0       &  \ldots     &  0        & 0       \\
  D_1    & 0          &        D_2     &  \ldots     &  0        & 0       \\
  0      & D_2        &        0       &  \ldots     &  0        & 0       \\
  \vdots &   \vdots   &    \vdots      &  \vdots     &  \vdots   & \vdots  \\
  0      &     0      &        0       &  \ldots     &  0        & D_2     \\
  0      &     0      &        0       &  \ldots     &  D_2      & 0
\end{array}
\right].
\eea
Diagonalization of the matrix
$D + 2\Omega$ is performed
by the unitary transformation
\bea{6}
D + 2 \Omega =  U \Lambda U^{+}, \quad
\Lambda = \mbox{diag}\{\lambda_1,\ldots ,\lambda_N \},
\eea
so that the new fermion operators
$\gamma^{+}_k $ and $\gamma_k$
introduced by the relations
\bea{7}
c_n^{+} =
 \sum\limits_{n=1}^{N} u_{n,k}^{*}\gamma_k^{+}
, \quad
c_n =
 \sum\limits_{n=1}^{N} u_{n,k}\gamma_k
\eea
bring the Hamiltonian \refeq{5} into the Hamiltonian
\bea{8}
H =
  \frac{1}{2}\sum\limits_{k=1}^{N} \lambda_k \gamma^{+}_k\gamma_k
- \frac{1}{2} \sum\limits_{n=1}^{N} \omega_n
\eea
with energies $\frac{1}{2}\lambda_{\nu}$
of the free fermion waves.

To go further in explicit calculations it is necessary to find the eigenvalues
$\lambda_{\nu}$
and eigenvectors
$\vert u_{\nu} \rangle =(u_{1\nu}, u_{2\nu},\ldots,
                            u_{N\nu})^{t}$
of the matrix
$D+2\Omega$,
\bea{9}
(D+2\Omega) \vert u_{\nu} \rangle =
\lambda_{\nu} \vert u_{\nu} \rangle.
\eea
In Eq. \refeq{9}, the components
$u_{ n,\nu}$ at the even sites $n=2,4,\ldots,N-1$
obey the equations
\bea{10}
D_1 u_{2k-1,\nu} + 2\omega_2 u_{2k,\nu} + D_2 u_{2k+1,\nu} = \lambda_{\nu}
u_{2k,\nu},
\quad k=1,2,\ldots,\frac{N-1}{2}.
\eea
We now fix the indices $k$ and $\nu$
in Eq. \refeq{10} and eliminate from Eq. \refeq{10} the components
$u_{2k-1,\nu}$ and  $u_{2k+1,\nu}$ at the odd sites
$(2k-1)$ and $(2k+1)$ by writing down the equations
for the $u_{2k-1,\nu}$, $u_{2k+1,\nu}$
from  Eq. \refeq{9} as
\bea{11}
D_2 u_{2k-2,\nu} +  2\omega_1 u_{2k-1,\nu} +  D_1 u_{2k,\nu}
& = & \lambda_{\nu} u_{2k-1,\nu} \nonumber   \\
D_2 u_{2k,\nu} +    2\omega_1 u_{2k+1,\nu} +  D_1 u_{2k+2,\nu}
& = & \lambda_{\nu} u_{2k+1,\nu}.
\eea
Substituting  \refeq{11} into  \refeq{10},
we get the relations for the amplitudes $u_{n,\nu}$ at the even sites,
\bea{12}
g u_{2k-2,\nu} + g u_{2k+2,\nu} =  u_{2k,\nu},
\quad k=1,2,\ldots,(N-1)/2,
\eea
with the spatially independent coupling constant
\bea{13}
g  =  \frac{ D_1 D_2 }
{ (\lambda_{\nu} - 2\omega_1)(\lambda_{\nu} - 2\omega_2) - D_1^2 - D_2^2 }.
\eea
The derivation of  \refeq{12}, \refeq{13} is a characteristic way
in the theory of the real space renormalization \cite{RG}
which eliminates the half of the degrees of  freedom belonging to the odd sites
yielding the field equations on the (even) lattice sites with doubled lattice constant.
Eq-s. \refeq{12}, \refeq{13} govern the amplitudes
$u_{2,\nu}, u_{4,\nu}, \ldots, u_{N-1,\nu}$
( including the amplitudes $u_{2,\nu}$, $u_{N-1,\nu}$
at the border sites $2$ and $N-1$ )
if we introduce the additional sites
$n=0$ and $n=N+1$ and put there
\bea{14}
u_{0,\nu} =0, \quad  u_{N+1,\nu} =0.
\eea
Eq. \refeq{14}
implies the cutting off the lattice at the sites $1$ and $N$, thus, preventing
the fermions from escaping the lattice shown on Fig. $1$.
The solution of
Eq. \refeq{12} conditioned by Eq. \refeq{14} read
\bea{15}
u_{2k,\nu} =  A_{\nu}\sin\left(
                       \frac{2 \pi k \nu}{N+1}
                \right), \quad
k, \nu =1,2,\ldots,(N-1)/2,
\eea
with eigenvalues
\bea{16}
\lambda_{\nu}^{(\pm)} & = &
 \omega_1 + \omega_2 \pm
\sqrt{
\left( \omega_1 - \omega_2 \right)^2 +
D_1^2 \: \Delta_{\nu}
     }, \quad  \nu=1,2,\ldots,\frac{N-1}{2},   \nonumber \\
\Delta_{\nu}       & = &
1   + 2\delta \cos
              \left(\frac{2 \pi \nu}{N+1}
              \right) + \delta^2 , \quad \delta = D_2/D_1.
\eea
Eq. \refeq{16} gives $2\frac{N-1}{2}= N-1$ eigenvalues
$\lambda_{\nu}$
since for each index $\nu = 1,2,\ldots,N $
and for each superscript $(+)$ and $(-)$
the relationship
$\lambda_{\nu}^{(\pm)}= \lambda_{N+1-\nu}^{(\pm)}$ holds.
It is convenient to arrange $(N-1)$ distinct eigenvalues
$\lambda_{\nu}$  as
\be{16.6}
\lambda_{\nu} = \cases{
\lambda_{\nu}^{(+)}, &  for $\nu=1,2,\ldots,(N-1)/2      $   \cr
\lambda_{\nu}^{(-)}, &  for $\nu=(N+3)/2,(N+5)/2,\ldots,N$   \cr
                      }
                              \quad \mbox{.}
\ee
In accord with the enumeration of the eigenvalues $\lambda_{\nu}$
in Eq. \refeq{16.6},
the missed  $N$-th  eigenvalue $\lambda_{\nu}$
stands for the index  $\nu = \frac{N-1}{2} + 1 = \frac{N+1}{2}$
(recall that $N$ is odd). To find the eigenvalue
$\lambda_{\frac{N+1}{2}}$
and the eigenvector
$u_{n,\frac{N+1}{2}}$ at even sites $n$, the use is made of the
properties
(see Eq. \refeq{15})
\bea{18}
u_{n,\frac{N+1}{2}} =  0, \quad n=0,2,4,\ldots,N-1,N+1.
\eea
By the condition  $u_{N-1,\frac{N+1}{2}} =  0$, the  $N$-th equation from  Eq. \refeq{9}
gives immediately the sought eigenvalue
\bea{19}
\lambda_{\frac{N+1}{2}} =
2\omega_1.
\eea
It remains to find the amplitudes $u_{k,\frac{N+1}{2}}$
at the odd sites. The sought amplitudes obey the relations
\bea{20}
D_1 u_{2k-1,\nu} + D_2 u_{2k+1,\nu} = 0, \quad
\nu=\frac{N+1}{2}, \: k=1,2,\ldots,\frac{N-1}{2}.
\eea
Eq. \refeq{20} gives the components of the eigenvector  $u_{n,\frac{N+1}{2}}$
belonging to odd sites $n$
(up to the normalization coefficient $B$),
\bea{21}
u_{n,\frac{N+1}{2}} = B \cdot (-\delta)^{ \frac{N-k}{2} },
 \quad n=1,3,\ldots,N.
\eea
Calculating the normalization coefficients
$A_{\nu}$ in Eq. \refeq{15} and the coefficient $B$
in Eq. \refeq{21}, one finds all
$N$  eigenvectors $\vert u_{\nu} \rangle$
and all
$N$  eigenvalues  $\lambda_{\nu}$ of the Hamiltonian \refeq{8},
\be{22}
\lambda_{\nu} = \cases{
                \omega_1 + \omega_2  +
                         \sqrt{
               \left( \omega_1 - \omega_2 \right)^2 +
    D_1^2 \:  \Delta_{\nu}      }, &   $\nu=1,2,\ldots,\frac{N-1}{2} $   \cr
                2\omega_1     , &   $\nu=\frac{N+1}{2} $   \cr
                \omega_1 + \omega_2  -
                        \sqrt{
               \left( \omega_1 - \omega_2 \right)^2 +
D_1^2  \:  \Delta_{\nu}
                              }, & $\nu=\frac{N+3}{2},\frac{N+5}{2},\ldots,N$ \cr
                      }
   \quad \mbox{.}
\ee
For all indices $\nu = 1,\ldots,N $ except the index  $\nu = \frac{N+1}{2}$,
the eigenvector
$\vert u_{\nu} \rangle$ has the elements
\be{22.2}
u_{j,\nu} = \cases{
                A_{\nu}\frac{D_1}{\lambda_{\nu} - 2\omega_1}
                        \left[
\delta \sin \left(\frac{ \pi \nu (j-1)}{N+1}\right) +
       \sin \left(\frac{ \pi \nu (j+1)}{N+1}\right)
                        \right]
                             , &  $j=1,3,5,\ldots,N $   \cr
                A_{\nu} \sin \left(\frac{ \pi \nu j}{N+1}\right)
                             , &  $j=2,4,\ldots,N-1 $ \cr
                      }
          \quad \mbox{.}
\ee
and the normalization coefficient
\bea{23}
A_{\nu} =
\frac{2 \vert \lambda_{\nu} - 2\omega_1 \vert}{\sqrt{N+1} }
\frac{1}
     { \sqrt{ (\lambda_{\nu} - 2\omega_1 )^2 +
        D_1^2 \:  \Delta_{\nu}
            }
     }.
\eea
By Eq. \refeq{21}, the elements
of the eigenvector
$\vert u_{ \frac{N+1}{2} } \rangle$ read
\be{24}
u_{j,\frac{N+1}{2}} = \cases{
                B\cdot(-\delta)^{ \frac{N-j}{2} }
                             , &  $j=1,3,5,\ldots,N $ \cr
                0
                             , &  $j=2,4,\ldots,N-1 $ \cr
                  }
   \quad \mbox{.}
\ee
with the normalization coefficient
\bea{25}
B = \biggl(
\frac{\delta^2 -1}{\delta^{N+1} - 1 }
    \biggr)^{\frac{1}{2}}.
\eea

%%%%%%%%%%%%%%%%%%%%%%%%%%%%%%%%%%%%%%%%%%%%%%%%%%%%%%%%%%%%%%%%%%%%%%%%%%%%%

\section{Spin-wave propagation in open chains
with alternating couplings
        }

\begin{figure}[ht]
\begin{center}
\includegraphics[scale=0.5]{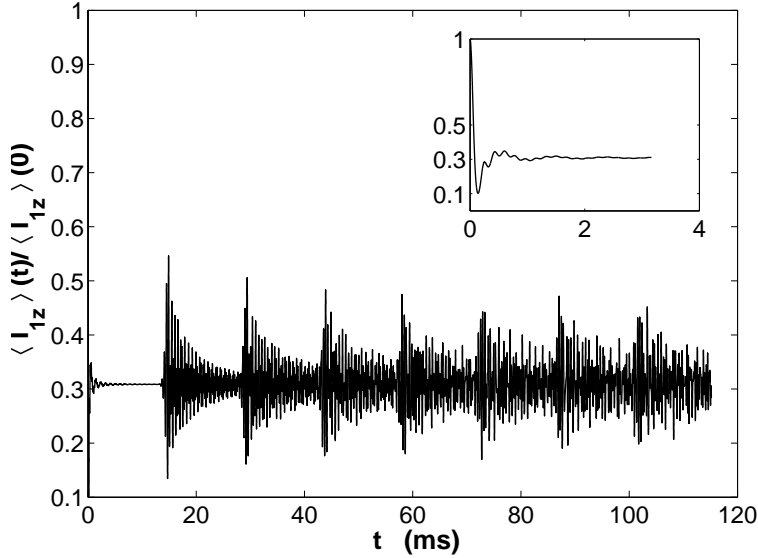}
\caption{
\small
\label{Fig2}
Time course of the polarization
$
\langle I_{1,z}\rangle (t)/ \langle I_{1,z}\rangle (0)
$
of the first spin, Eq. \refeq{29},
for
the chain of $201$ spins interacting with coupling coefficients
$D_1= 2\pi\cdot 4444 \: s^{-1}$
and $D_2= 2\pi\cdot 6666 \: s^{-1}$ and equal Larmor frequencies.
The initial polarization is on the first spin. Insert shows the
time course of the polarization on an earlier time interval
$0 \le t < 4$ ms.
     }
\end{center}
\end{figure}

The spectrum of Eq.-s \refeq{22} - \refeq{25} of the $XY$ Hamiltonian
 \refeq{8}
can now be applied to describe how
the polarization  at a single site of the alternating chain with
equal Larmor frequencies $\omega_n$
changes over the time.

Let the initial polarization is on the single site
$j$, hence, the spin dependent part of
the initial density matrix
at the high temperature
approximation is, see \cite{Abragam},
\bea{26}
\rho(0) = I_{j,z} =  c_j^{+}c_j - \frac{1}{2}.
\eea
Given
the initial density matrix \refeq{26} and
the Hamiltonian $H$ \refeq{8},
the Liouville-von Neumann equation ($\hbar = 1$)
\bea{27}
 i \frac{\partial \rho}{\partial t}  =
 \Bigl[ H, \rho
 \Bigr],
\eea
is solved
in terms of the fermion operators \refeq{7} as follows,
\bea{28}
\rho( t) & = & e^{-iHt} I_{j,z} e^{iHt} = \nonumber \\
         & = & -\frac{1}{2} + \sum_{1 \le l,m \le N} u^{*}_{j,l}u_{j,m}
               e^{-\frac{i}{2}(\lambda_l-\lambda_m)t }
           \gamma^{+}_l \gamma_m .
\eea
Denoting the polarization at $j'$-th spin at the time moment $t$
by
$\langle I_{j',z}\rangle (t)$ and invoking  Eq. \refeq{26}, we get
\bea{29}
\frac{\langle I_{j',z}\rangle (t) }
     {\langle I_{j,z}\rangle (0)} =
\frac{ tr\{ \rho(t) I_{j',z}   \} }
     { tr\{         I_{j,z}^2 \} } =
\biggl |
         \sum_{1 \le \nu \le N} u^{*}_{j,\nu}u_{j',\nu}
               e^{-\frac{i}{2}\lambda_{\nu}t }
\biggr |^2 .
\eea
Fig. $2$ shows the time course of the polarization
$
\langle I_{1,z}\rangle (t)/ \langle I_{1,z}\rangle (0)
$
at site $j'=1$
when the initial polarization is also at  site $1$ and
$D_2 = \frac{3}{2} D_1$.
Just as in the case of the polarization dynamics on the homogeneous chain with
equal couplings,
$ D_2 = D_1 $,
\cite{FBE}, the dynamics of the polarization of the alternating chain can be regarded as the
propagation of the spin wave packet starting at site $1$
and bouncing back and forth at the chain ends
To calculate the return time $t_{\sf R}$ for the wave packet to reappear
at site $1$, we, first, determine
the group velocity of the waves described by the Hamiltonian $H$ \refeq{8}
with the dispersion law
$\frac{1}{2}\lambda_{\nu}$ \refeq{16} and equal Larmor frequencies, $\omega_1 = \omega_2$.
By specifying the wave vector
$p=2\pi \nu/(N+1)$, $\nu = 1,2,\ldots,(N-1)/2$, the
dispersion law of Eq. \refeq{16} written down as
$   \: \:
\frac{1}{2} \lambda_{\nu} =
\frac{1}{2} D_1 \sqrt{
       1   + 2\delta \cos(p) + \delta^2
         } \:  \:
$
 allows one to calculate the sought group velocity
\bea{29.1}
\upsilon = \mathop{\max}\limits_p
           \left\{
            a \: \frac{d\lambda(p)}{2\:dp}
           \right\} = \frac{a D_1}{2} ,
\eea
where $a$ is the lattice constant
and the coupling  constant $D_1$ is the minimal coupling constant
among the two coupling constants  $D_1$ and $D_2$.
Thus, for
$N=201$-chain with coupling constants
$D_1= 2\pi\cdot 4444 \: s^{-1}$
and  $D_2= 2\pi\cdot 6666 \: s^{-1}$,
the time of the first returning of the wave packet to site $1$
becomes,
see Fig. $2$,
\bea{30}
t_{\sf R} =\frac{2(N-1)}{D_1} \approx 14.5 \: ms,
\eea
The traveling waves of the spin polarization
$
\langle I_{j,z}\rangle (t)/
\langle I_{1,z}\rangle (0)
$
are shown on Fig. $3$.

\begin{figure}[ht]
\begin{center}
\includegraphics[scale=0.7]{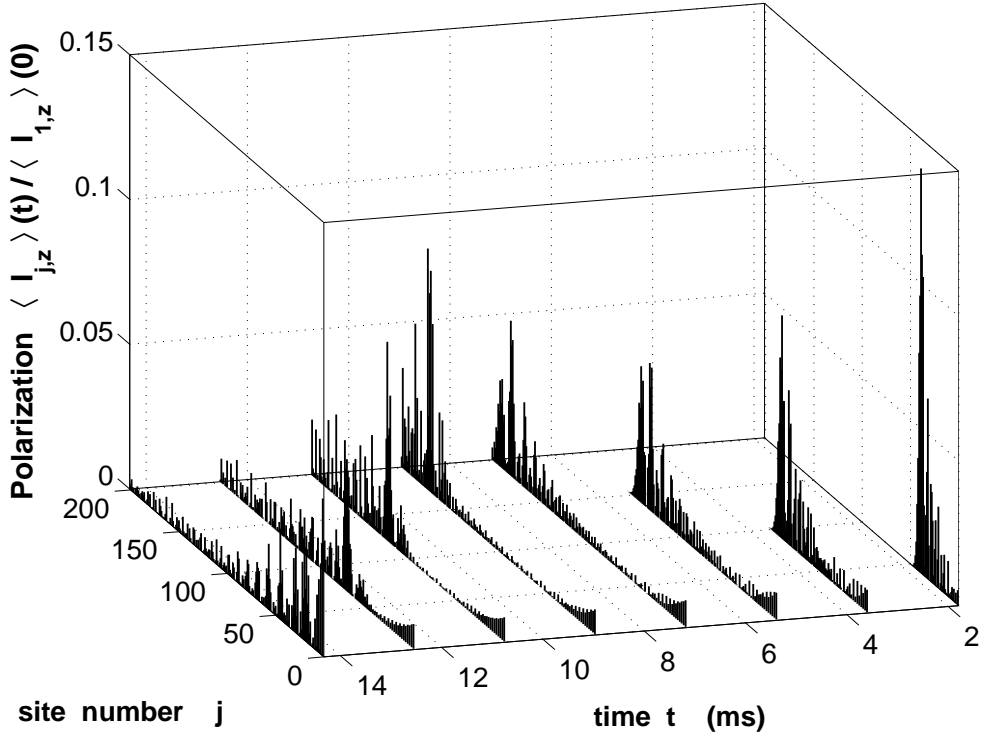}
\caption{
\small
\label{Fig3}
Propagation of the spin-wave packet  along the $201$-site open chain.
The packet starts at boundary site $j=1$.
     }
\end{center}
\end{figure}

\section{
Intensities of MQ coherences in open chains
with alternating couplings
        }

The exact spectrum
of Eq.-s \refeq{22} - \refeq{25} of the Hamiltonian  $H$ \refeq{8}
provides us with the technique for determining of the multi-quantum
dynamics in an alternating open chain.
Again, we take the initial density matrix
$\rho(0)$ \refeq{26} and calculate how
the
MQ coherences  develop in the spin system of the
alternating spin-$\frac{1}{2}$ chain. MQ NMR dynamics of the
nuclear spins coupled by nearest neighbour dipolar interactions are described by
the Hamiltonian \cite{BMGP}, \cite{DMF},
\bea{31}
H_{\sf MQ} = \frac{1}{2}
\sum\limits_{ n=1}^{N-1}
D_{n,n+1} \left\{
I_{n,+} I_{n+1,+} + I_{n,-} I_{n+1,-}
\right\}.
\eea
The  Hamiltonian $H$ \refeq{31} takes the form of the exactly solvable Hamiltonian $H$ \refeq{1}
(with the Larmor frequencies $\omega_n=0$ for all sites) by making use
of
the unitary transformation \cite{DMF} acting on the even sites,
\bea{32}
Y = \exp(-i\pi I_{2,x}) \exp(-i\pi I_{4,x}) \cdots \exp(-i\pi
I_{N-1,x}),
\eea
so that $Y H_{\sf MQ}Y^{+} = H (\mbox{ Eq.}\refeq{1}; \{ \omega_n=0 \} )$.
In addition, the transformation  $Y$
brings the initial density matrix \refeq{26} to the form
\bea{33}
\bar \rho(0) = Y I_z Y^{+} = \sum_{n=1}^{N} (-1)^{n-1} I_{n,z},
\eea
where we introduce the total polarization
$
I_z = \sum\nolimits_{n=1}^{N} I_{n,z}
$.
The Liouville-von Neumann Eq.
\refeq{27} with
Hamiltonian $H$ \refeq{31} and the initial density matrix
\refeq{26}
gives the intensities $G_n(t)$ of $n=0$ and $n=\pm 2$ orders,
just as in the case of the homogeneous chain
with the conservation condition \cite{DMF}, \cite{LHGM},
\bea{35}
G_0(t) +  G_2(t) + G_{-2}(t) =1,
\eea
and
\bea{36}
G_{0}(t)=
\frac{1}{N}
         \sum_{n=1}^{ N } \cos^2(\lambda_{\nu} t ), \quad
G_{\pm 2}(t)=
\frac{1}{2N}
         \sum_{n=1}^{ N } \sin^2(\lambda_{\nu} t ).
\eea
Fig. $4$ demonstrates the development of the $2$Q coherence
on the alternating  $N=201$-chain with couplings
$D_1= 2\pi\cdot 4444 \: s^{-1}$
and  $D_2= 2\pi\cdot 6666 \: s^{-1}$, thus $\delta = \frac{D_2}{D_1} = 1.5$.
As time proceeds, the regular course of the intensities $G_{\pm2}(t)$
is transformed to the erratic temporary behaviour just as in the
case of the homogeneous lattice with $D_1 = D_2$ \cite{DFL}.
For $D_1 = D_2$, Eq. \refeq{36} reproduces exactly the results for
the intensities of the $2$Q coherence of Ref. \cite{DMF}.

\begin{figure}[ht]
\begin{center}
\includegraphics[scale=0.6]{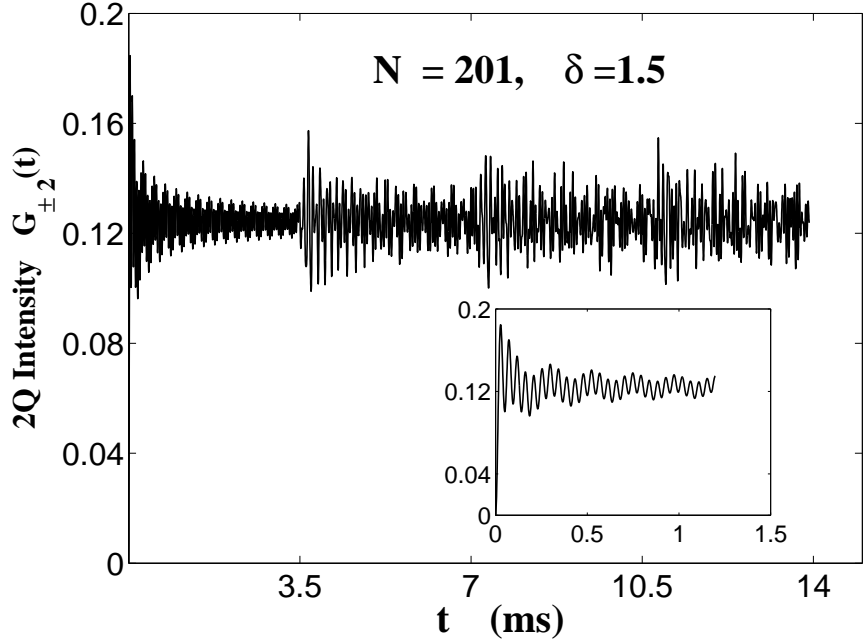}
\caption{
\small
\label{Fig4}
Time course of the intensities of the $2$Q coherence, Eq. \refeq{36}, in
the chain of $N=201$ spins interacting with coupling constants
$D_1= 2\pi\cdot 4444 \: s^{-1}$
and $D_2= 2\pi\cdot 6666 \: s^{-1}$ and zero Larmor frequencies.
Insert shows an early development of $2$Q coherence over
times $t < 1.5$ ms.
       }
\end{center}
\end{figure}

%%%%%%%%%%%%%%%%%%%%%%%%%%%%%%%%%%%%%%%%%%%%%%%%%%%%%%%%%%%%%%%%%%%%%%%%%%%%%

\noindent
\section{Conclusion}
\label{sec. 4}

Sensitivity of the NMR spin polarization dynamics to the spatially
periodic short-distance inhomogeneity of the lattice has been
explored in the previous sections relying on the exact spectrum of
the spin-$\frac{1}{2}$ $XY$-Hamiltonian
on the alternating open chain with an odd number of sites.

If the spin polarization is prepared initially on the single site, the
spin dynamics on the non-homogeneous lattice is represented as the
propagation of the spin wave packet. The velocity of the wave
packet on the non-homogeneous lattice is controlled by the minimal
coupling constant $\min\{D_1, D_2\}$. In the extreme situation
$D_1 \ll D_2$ ($D_1, D_2 \ne 0$), the spin polarization stays
fixed on the initial site.

As time proceeds, the regular spin
propagation along the alternating chain is transformed to the
erratic one.

The time scale of the regular behaviour
of the $2Q$ intensities $G_{\pm2}(t)$ is
$4$ times shorter  than
the time scale of the regular behaviour
of the polarization
$
\langle I_{j,z}\rangle (t)/
\langle I_{1,z}\rangle (0)
$,
as it can be seen by a comparison of Fig. $4$ with Fig. $2$.
The same effect happens in the case of the spin dynamics on the
homogeneous chain \cite{DFL}. The effect is caused by  two reasons.
First, the velocity of the propagation of the $2Q$ coherence is doubled  as
compared with  the velocity of the propagation of the spin polarization, as
it is obvious by the comparison  of the dispersion law in Eq. \refeq{36}
with
the dispersion law in Eq. \refeq{29}. Secondly, the two-spin
local excitations of the $2Q$ coherence travel the path $N$ before the two-spin
excitations return back on the $N$-site chain. The path $N$ should be compared
with the path $2N$ traveled  by the single
local spin excitation before it reappears on the initial site of the chain.

\vskip 5mm

Authors are thanking
A.A. Belavin, O.V. Derzhko, S.I. Doronin, A.I. Smirnov and M.A. Yurishev
for the discussions of the problems involved as well as thank to I.I. Maksimov for the
support in preparing the manuscript. The work is
supported through Grant RFBR $04-03-32528$.

%%%%%%%%%%%%%%%%%%%%%%%%%%%%%%%%%%%%%%%%%%%%%%%%%%%%%%%%%%%%%%%%%%%%%%%%%%

%\newpage  % FOR FIGURE

%%%%%%%%%%%%%%%%%%%%%%%%%%%%%%%%%%%%%%%%%%%%%%%%%%%%%%%%%%%%%%%%%%%%%%%%%%

%\newpage

% begin{figure}%[h] %h=here; b=bottom; p=separate page;t=top
% \begin{picture}(10.,10.)
% \includegraphics*[angle=0,width=13cm,height=10cm]{test7.eps}
% \end{picture}
% \end{figure}

%%%%%%%%%%%%%%%%%%%%%%%%%%%%%%%%%%%%%%%%%%%%%%%%%%%%%%%%%%%%%%%%%%%%%%%%%%
%%%%%%%%%%%%%%%%%%%%%%%%%%%%%%%%%%%%%%%%%%%%%%%%%%%%%%%%%%%%%%%%%%%%%%%%%%

% Fig. $1$.
% Open chain of odd number, $N$, of spins
% with the alternating NN couplings  $D_1$ and $D_2$
% and alternating Larmor frequencies $\omega_1$ è $\omega_2$.

% Fig. $2$.
% The time course of the polarization
% $
% \langle I_{1,z}\rangle (t)/ \langle I_{1,z}\rangle (0)
% $
% of the first spin, Eq. \refeq{29},
% for
% the chain of $201$ spins interacting with couplings
% $D_1= 2\pi\cdot 4444 \: s^{-1}$
% and $D_2= 2\pi\cdot 6666 \: s^{-1}$ and zero Larmor frequencies.
% Initial polarization is on the first spin.
% The polarization state at the time moment of the first returning
% of the spin packet to the first site is shown by arrow.

% Oscillations of the $2$Q coherences, Eq. \refeq{36}, in
% the chain of $N=201$ spins interacting with couplings
% $D_1= 2\pi\cdot 4444 \: s^{-1}$
% and $D_2= 2\pi\cdot 6666 \: s^{-1}$ and zero Larmor frequencies.
% Insert shows an early development of $2$Q coherences over
% times $t$ obeying $0\le tD_1 \le 20$.

%%%%%%%%%%%%%%%%%%%%%%%%%%%%%%%%%%%%%%%%%%%%%%%%%%%%%%%%%%%%%%%%%%%%%%%%%%

\newpage
\begin{figure}
%[h] %h=here; b=bottom; p=separate page;t=top
%\includegraphics*[angle=90,width=13cm,height=10cm]{chain_22.eps}
%\includegraphics[angle=270,scale=0.5]{chain22.eps}
% \caption{}
\end{figure}

%%%%%%%%%%%%%%%%%%%%%%%%%%%%%%%%%%%%%%%%%%%%%%%%%%%%%%%%%%%%%%%%%%%%%%%%%%
%%%%%%%%%%%%%%%%%%%%%%%%%%%%%%%%%%%%%%%%%%%%%%%%%%%%%%%%%%%%%%%%%%%%%%%%%%

\newpage
% \includegraphics[scale=1]{chain25.eps}
% \includegraphics[scale=0.5]{chain25.eps}

%%%%%%%%%%%%%%%%%%%%%%%%%%%%%%%%%%%%%%%%%%%%%%%%%%%%%%%%%%%%%%%%%%%%%%%%%%
%%%%%%%%%%%%%%%%%%%%%%%%%%%%%%%%%%%%%%%%%%%%%%%%%%%%%%%%%%%%%%%%%%%%%%%%%%

%\newpage
%\includegraphics[scale=0.5]{polarization_2.eps}

%%%%%%%%%%%%%%%%%%%%%%%%%%%%%%%%%%%%%%%%%%%%%%%%%%%%%%%%%%%%%%%%%%%%%%%%%%


\begin{thebibliography}{99}
\bibitem{LSM} E. Lieb, T. Schultz, D. Mattis,
              Ann. Phys. (N.Y.) {\bf 16}, 407,  (1969).
\bibitem{CG}  H.B. Cruz, L.L. Gonsalves,
              J. Phys. C {\bf 14}, 2785, (1981).
\bibitem{FBE} E.B. Fel'dman, R. Br\"uschweiler, R.R. Ernst,
              Chem.Phys.Lett. {\bf 294}, 297, (1998).
\bibitem{MBSBE} Z.L. Madi, B. Brutscher, T. Schulte-Herbr\"uggen,
                R. Br\"uschweiler, R.R. Ernst,
                Chem.Phys.Lett. {\bf 268}, 300, (1997).
\bibitem{PL} H.M. Pastawski, P.R. Levstein, G. Usaj,
                  Phys. Rev. Lett. {\bf 75}, 4310, (1995).
\bibitem{DR} O. Derzhko, J. Richter,
              Phys. Rev. {\bf 55}, 14298, (1997).
%\bibitem{DPL} E.P. Danieli, H.M. Pastawski, P.R. Levstein,
%              Physica {\bf B 320}, 351, (2002).
\bibitem{YDX} F. Ye, G.-H. Ding, B.-W. Xu,
              ArXive {\sf e-print}  cond-mat/0105584.
\bibitem{RG}  J. Machta, Phys. Rev. {\bf B 24} 5260, (1981);\\
              R.B. Stinchcomb, J.Phys. {\bf A 18}, L 591,(1985).
\bibitem{Abragam} A. Abragam, The Principles of Nuclear Magnetism.
                              Oxford, Clarendon Press (1961).
\bibitem{BMGP} J.Baum, M. Munowitz, A.N. Garroway, A. Pines,
                 J. Chem. Phys. {\bf 83}, 2015, (1985).
\bibitem{DMF} S.I. Doronin, I.I. Maksimov, E.B. Fel'dman,
              JETP {\bf 91}, 597, (2000).
\bibitem{DFL} S.I. Doronin, E.B. Fel'dman, S. Lacelle,
              Chem.Phys.Lett. {\bf 353}, 226, (2002).
\bibitem{LHGM} D.A. Lathrop, E.S. Handy, K.K. Gleason,
               J. Magn. Reson., {\bf A 111}, 161, (1994).

\end{thebibliography}
\end{document}